\documentclass{article} 
\usepackage{digital_commons,times}

\usepackage{hyperref}
\usepackage{url}
\usepackage{tabularx}

\title{Generative AI and the Digital Commons}

\author{Saffron Huang \\
Collective Intelligence Project \\
\texttt{saffron@cip.org} \\
\And
Divya Siddarth \\
Collective Intelligence Project \\
\texttt{divya@cip.org} \\
}

\iclrfinalcopy 
\begin{document}

\maketitle

\begin{abstract}
Many generative foundation models (or GFMs) are trained on publicly available data and use public infrastructure, but 1) may degrade the “digital commons” that they depend on, and 2) do not have processes in place to return value captured to data producers and stakeholders. Existing conceptions of data rights and protection (focusing largely on individually-owned data and associated privacy concerns) and copyright or licensing-based models offer some instructive priors, but are ill-suited for the issues that may arise from models trained on commons-based data. We outline the risks posed by GFMs and why they are relevant to the digital commons, and propose numerous governance-based solutions that include investments in standardized dataset/model disclosure and other kinds of transparency when it comes to generative models’ training and capabilities, consortia-based funding for monitoring/standards/auditing organizations, requirements or norms for GFM companies to contribute high quality data to the commons, and structures for shared ownership based on individual or community provision of fine-tuning data.
\end{abstract}

\section{The rise of Generative Foundation Models (GFMs)}

We will use the phrase “generative foundation models” (GFMs) to refer to machine learning systems that are: 1) “generative” — they generate text, images, or other sequences of information based on some input prompt, and 2) “foundation models” — neural network models trained on a large dataset comprising diverse origins and content, and can be adapted to a wide range of tasks. (Machine learning, or ML, is sometimes also referred to as artificial intelligence, or AI).
Examples of well-known GFMs are: OpenAI’s GPT family of language models (including ChatGPT) that take in text and generate text; DALLE-2, which takes in text/images and generates images; BERT, which takes in text and generates text, Stable Diffusion, which takes in text and generates images; Codex, which takes in code (a specific kind of text) and generates code.

We speak of “generative” foundation models, rather than foundation models at large, per \citet{bommasani2022opportunities}, because  we are concerned primarily with the applicability of these models for generating content, such as generating text, code or images. This may include tasks such as summarization (generating a summary of a text) or text continuation (continuing the text by iteratively predicting the next word) or creating images and videos.
 
Generative foundation models are a general technology, although they benefit from adaptation to the “downstream” tasks they are used for, e.g. by “fine-tuning” them by training on more specific datasets such that they can generate the appropriate material for the use context.
The structure of the nascent industry is likely to greatly change, but at the moment there are a few key actors creating more general-purpose GFMs, such as OpenAI, Midjourney, EleutherAI, BigScience, and Stability. More-specialized companies often build off the technology released by the actors above, applying the technology for specific tasks such as:
\begin{itemize}
    \item Copywriting (e.g. \href{https://www.copy.ai/}{Copy.ai}, \href{https://jasper.ai/}{Jasper}, \href{https://www.neuraltext.com/}{NeuralText}, \href{https://nichesss.com/}{Nichesss})
    \item Website generation (e.g. \href{https://www.the.com/}{The.com}, \href{https://debuild.app/}{Debuild})
    \item Marketing and stock image generation (e.g. \href{https://www.theverge.com/2022/10/25/23422359/shutterstock-ai-generated-art-openai-dall-e-partnership-contributors-fund-reimbursement}{Shutterstock}, \href{https://picsart.com/}{Picsart}, \href{https://www.jasper.ai/art}{Jasper Art})
    \item General image editing and design (e.g. \href{https://medium.com/codex/stable-diffusion-arrives-in-photoshop-heres-how-to-install-3db277491023}{Photoshop plugin}, \href{https://techcrunch.com/2022/10/12/microsoft-brings-dall-e-2-to-the-masses-with-designer-and-image-creator/}{Microsoft’s Designer})
    \item General writing, editing, and content management (e.g. \href{https://www.microsoft.com/en-us/microsoft-365/blog/2023/02/01/microsoft-teams-premium-cut-costs-and-add-ai-powered-productivity/}{Microsoft Teams Premium}, \href{https://www.notion.so/product/ai}{NotionAI}, \href{https://coda.io/@coda/openai-for-coda}{Coda +
OpenAI}, \href{https://chibi.ai/}{Chibi AI})
    \item Video editing (e.g. \href{https://runwayml.com/}{Runway}, \href{https://trykino.com/}{Kino})
    \item Code generation (e.g. \href{https://github.com/features/copilot}{Github Copilot}, \href{https://www.tabnfine.com/}{Tabnine})
    \item Research assistants (e.g. \href{https://ought.org/elicit}{Elicit})
    \item Building floor plans for home renovations (e.g. \href{https://www.tailorbird.us/}{Tailorbird})
    \item Excel/data wrangling and cleaning (e.g. \href{https://charm.shubhro.com/}{Charm})
\end{itemize}
Note that this is only a subset of the current and possible GFM applications, particularly as the industry is moving rapidly.

\section{The promise and peril of GFMs}

New general-purpose technologies come with both risks and opportunities. Over the next decade, GFMs may contribute to enormous economic wealth and productivity, underpinning advancements across sectors—from lightning-fast copywriting to accessible legal services to automated (and increasingly, autonomous) online interactions. Furthermore, while the initial crop of GFMs was put forward by a small number of firms, the recent proliferation of open-source models and applications may assuage initial fears regarding the hyper-concentration of this productive capacity, pointing towards a potentially more open and competitive ecosystem.
However, there is no guarantee that GFMs will be deployed for the broad benefit of humanity. The growth and expansion of these technologies carry with them risks that need to be investigated and mitigated where appropriate, including:

\begin{enumerate}
    \item \textbf{Poisoning the information sphere with easy-to-create low-quality data.} GFMs can generate text/code/images much faster than humans, but can output untrue, biased or otherwise low-quality material. This may lead to a proliferation of both accidental and deliberate mis/dis-information and biased, inappropriate, or low-value content. Research has shown that stereotypes tend to be amplified through text-to-image GFMs \citep{bianchi2022easily}. Outputs may also be subtly incorrect and difficult to evaluate, making quality assurance a difficult task without dedicated auditing capabilities.
    \item \textbf{Eroding self-determination and democracy.} GFMs could be used for personalized persuasion or disinformation, churning out harmful information at low cost \citep{goldstein2023generative}. As GFMs are used in a greater number of applications, their black-box decision-making may affect material outcomes for many. Content filters and other design decisions are generally determined by a small team with little participation or calibrated input, even if models are later open-sourced \footnote{This can lead to issues, such as with Meta’s Galactica AI filtering out legitimate scientific research to do with race or AIDs \citep{heaven2022meta}}.
    \item \textbf{Homogenizing content.} Generated information becomes more homogenous, with  similar properties (e.g. sentiments on popular issues) due to being based on one or a few widely-used models.
    \item \textbf{Misaligning incentives for humans to contribute to the open digital ecosystem.} People may stop producing text/code/images in favor of using generative models, or never visit the websites from which data is sourced (like Reddit or Wikipedia)  in favor of using GFMs as search engines, and thus lead to decreased contributions to them (a phenomenon which has been referred to as the “paradox of reuse”) \citep{mcmahon_johnson_hecht_2017} People may also decide against releasing non-GFM-enabled creations into the commons e.g. due to fears of labor replacement or lack of attribution.
    \item \textbf{Driving further economic concentration.} If certain capital-intensive models are privately owned with limited or no outside access (e.g. Google’s PaLM, DeepMind’s Gopher, OpenAI’s DALLE-2) or significant control over applications built on top, this could contribute to economic concentration. Access may be limited particularly if the high costs of creating GFMs don’t decrease soon.
    \item \textbf{Contributing to precarious labor conditions and large-scale automation.} For certain industries the automation of some or many parts of human work could contribute to precarious labor conditions and potentially lead to large-scale automation (e.g. digital/concept art, copywriting), issues which labor policy must then address.
    \item \textbf{Accelerating unpredictable risks from highly capable AI systems.} Black-box, highly capable artificial intelligence systems can be dangerous in unpredictable ways. GFMs are not necessarily built as autonomous agents, but others can add the ability to use software, traverse the internet or manipulate physical objects which can extend the possible domain of risks \citep{reed2022a}.
\end{enumerate}

Note that very few of these issues have been studied in the GFM context thus far, and none extensively. A large part of our proposed solutions will entail setting up the infrastructure to monitor and study these effects.

\section{GFMs and the commons}
\subsection{The core of many risks is the effect of GFMs on the digital commons.}
GFMs both depend on and contribute to what is often known as the “commons” \citep{ostrom1990governing}. Commons, more specifically common-pool resources, are resources that are both rival and non-excludable, and may thus fall prey to ‘tragedy of the commons’ style exploitation, in which individual actors can free-ride on, poison, or otherwise damage shared resources at great collective cost. Effective multi-stakeholder governance and management thus determines the sustainability of these common-pool resources.  
We limit analysis primarily to the \textit{digital} commons, as that is the realm in which GFMs are primarily created and deployed. The digital commons comprises two things:

\begin{enumerate}
    \item \textbf{The online commons of information resources that we all benefit from, own and contribute to together.} This traditionally includes things like wikis, Internet archive snapshots, Creative Commons (CC) licensed images and public software repositories, but online discussion spaces such as Reddit and news sources such as The Guardian are also part of this information commons. Much of this data or copies are technically hosted by someone (e.g. a private entity, the government, individuals) and authority/management over them vary in their level of collective control/ownership. But these generally have reasonably open and shared access, with few barriers to people contributing to or using these digital resources.
    \item \textbf{The collective infrastructure that underpins the commons.} This infrastructure includes the physical (e.g. Internet cables), the institutional (e.g. organizations like the IETF, W3C, and IEEE), and the technological (e.g. open-source libraries). As we see increasingly more GFMs and other AI being deployed, we may see more ML models, datasets, libraries and platforms, ML-tailored computing hardware, as well as various AI building or governing institutions come under this umbrella. For example, if GFM products begin to replace traditional software products, machine learning libraries such as Pytorch may become a critical piece of digital infrastructure supporting an increasing amount of online functionality.
\end{enumerate}

The existence and quality of the digital commons can be threatened by “undersupply, inadequate legal frameworks, pollution, lack of quality or findability” and needs to be maintained against such outcomes \citep{digital_commons} For example, people, such as coders, artists, Wikipedians and bloggers, contribute and maintain high quality material in our information commons. Spam filters keep undesirable solicitations out of inboxes, the Creative Commons non-profit provides a multiplicity of licenses to make flexible copyright terms possible, internet archivists keep information available and open-source contributors create tools to support much of the world’s software.

The idea of digital commons can lend well to ML governance in particular:

\begin{quotation}
Digital commons could also be an answer to the need for new governance structures for resources such as data or artificial intelligence… \textbf{Data lends itself especially well to a commons framework:} both inputs and impacts are fundamentally shared, distributing access to these resources provides a foundation for further bottom-up innovation and technological progress, siloing or privatizing these erodes the possibility of stewarding collective benefit… [forming] a shared layer necessary for economic growth and democratic participation. \citep{siddarth_weyl_2021}
\end{quotation}

With respect to each of the risks detailed above, risks 1)-4) are most straightforwardly related to the commons-based approach, as they are directly concerned with the impact on common resources and the public sphere. Risks 5)-6), concerned with economic concentration and labour precarity/automation, raise questions around what obligations are tied to common resources. In short, to what extent does public data generate obligations whereby the people who created it should receive value from the resulting AI? If the data is included in the production of GFMs used for private interests and against the interest of those very people that created it, questions of compensation and reward become more salient.

\subsection{The integrity of the digital commons matters.}

The digital commons enables broadly shared access and benefits from digital technologies. The economic and ethical benefits of open-source have been studied repeatedly \citep{ghosh2007economic, wright2021open, blind2021impact}, open access policies amplify the diffusion of science and Wikipedia is one of the most visited websites in the world \footnote{Wikipedia is more likely to reference open access and higher impact-factor journals according to \citet{teplitskiy2017amplifying}, and is also one of the most visited websites according to multiple analyses \citep{atlantic_lafrance, domantas}}. Because these are common resources and are not monetized, their value is difficult to measure, but accessibility seems clearly to lead to shared benefit, especially with non-rivalrous digital goods. Without shared collective infrastructure, it is possible that current and future innovations can be easily dominated by private entities. For example, in the 20th century, AT\&T gained a leading monopoly over cable and radio due to its existing infrastructural monopoly over long-distance telephone lines \citep{wu_2012}). A lack of collective digital infrastructure would potentially lead to monopoly dynamics rather than healthy competition that drives innovation.

The digital commons is critical to modern knowledge-sharing. Knowledge in general is also seen as a shared, common-pool resource, referred to as “the knowledge commons” and analyzed as such \citep{knowledge_commons} The digital commons, in this age, is a key part of the contemporary knowledge commons. Accurate, accessible, comprehensive and diverse sources of knowledge are widely acknowledged as necessary to culture, welfare, science and technology. Both the training data underlying GFMs and the outputs are part of this knowledge commons, and the GFM itself can be seen as a kind of interface to the knowledge in its training data.

The digital commons underpins democracy. High quality knowledge and genuine debate between humans is important in itself, but also to democracy. Good information and public discourse are necessary for making good decisions on who to vote for or what policies to enact, and for holding representatives accountable. As pointed out:

\begin{quotation}
    The foundational mechanism upon which all others depend is the maintenance of a healthy epistemic commons within a democracy—an \textbf{epistemically healthy public sphere where widely trusted norms, processes, and institutions for making sense out of and reaching consensus on raw information} lead to certain facts being accepted as true. \citep{consilience}
\end{quotation}

Nowadays, as many of these interactions move online, we have seen how degradation of the digital knowledge commons can harm democracy, e.g. through misinformation, polarization, and other means \cite{del2016spreading, barrett2021fueling, anderson2020concerns}.

\subsection{Generative foundation models rely on, but may also erode, the digital commons.}

\textbf{GFMs are trained on the digital commons.} Generative foundation models leverage large databases of scraped information (text, code, images) from the internet to train highly capable models. This depends on the availability of public, scrapable data and leverages the “collective intelligence” of humanity, including the painstakingly edited Wikipedia, millennia's worth of books, billions of Reddit comments, hundreds of terabytes’ worth of images, and more \footnote{LAION-5B, which Stable Diffusion is trained on, has 5 billion text-image pairs \citep{schuhmann2022laion}. The Pile has 100+GB of books \citep{gao2020pile}.}. They also rely on non-profits like Common Crawl (which build and maintain open repositories of web crawl data), Creative Commons (for open licenses for the data used), open source libraries, and other digital infrastructure. They also take advantage of aggregated user preferences; e.g. the WebText dataset underlying the GPT family of models uses Reddit “karma scores” to select content for inclusion. All of this is common digital information and infrastructure that many people contribute to.

\textbf{The dependence of GFMs on digital commons has economic implications}: much of  the value comes from the commons, but the profits of the models and their applications may be disproportionately captured by those creating GFMs and associated products, rather than going back into enriching the commons. Some of the trained models have been open-sourced, some are available through paid APIs (such as OpenAI’s GPT-3 and other models), but many are proprietary and commercialized. It is likely that users will capture economic surplus from using GFM products, and some of them will have contributed to the commons, but there is still a question of whether there are obligations to directly compensate either the commons or those who contributed to it.

\textbf{In addition, there are legal and moral implications. }Laws to do with copyright and fair use have little precedent when it comes to generative AI, but there are already multiple lawsuits challenging the use of public data in a variety of such GFMs, including Github Copilot, Stability AI and Midjourney \citep{lawsuits}. Some companies, such as DeviantArt, are training models on user images, requiring them to opt-out (rather than opt-in) to the training set. Furthermore, artists who have released work have unwittingly been subjected to others training image models on their creations and creating outputs that they feel are invasive/don’t reflect the style or intent of their actual work \citep{invasive_diffusion}.

\textbf{GFMs put material back into the digital commons much faster than humans can, and this material is of unknown quality.} As GFM applications proliferate, copywriting tools start to fill the internet with AI-generated marketing text, website copy and blog posts. Image generation tools are used for blog posts, presentations, and website decoration. Everyday people use language GFMs to help them write blog posts and articles. The large-scale characteristics of this text for many use cases is as yet unknown, and there are potential effects of large-scale biasing or misinformation. While some of the Internet is potentially already auto-generated to some extent, GFMs may be differentiated in their:

\begin{enumerate}
    \item \textit{Speed.} These generative models can generate text/code/images at much higher speed than humans can write/code/make art themselves. This machine-generated text could become a high \% of the digital information commons, and potentially also homogenize it (information starts to have similar properties as they come from similar models). One analysis estimates that we will run out of high-quality language data for ML training by 2026 \citep{epoch2022willwerunoutofmldataevidencefromprojectingdataset}.
    \item \textit{Quality.} 
    \begin{enumerate}
        \item GFMs often say untrue or biased things, and the incorrectness is often subtle rather than easy to catch and correct \citep{sobieszek2022playing} Generated language outputs are likely to over-predict rare rather than trivial events, a.k.a. reporting bias \citep{shwartz2020neural}, and replicate discriminative biases of humans \footnote{The Gopher paper shows that this 280bn-parameter generative language model exhibits gender and occupational bias, sentiment bias towards different social groups \citep{rae2022scaling}.}; Stable Diffusion has been found to replicate stereotypes, and Galactica (Meta’s science language model) was quickly taken down after outputting falsehoods and prejudiced statements that nevertheless sounded plausible \citep{stable_diff_bias, heaven2022meta}. Sometimes generations are more straightforwardly low-quality, being e.g. repetitive, unable to comprehend negations or unable to do common-sense inference \footnote{The repetition problem is covered in \citet{fu2021theoretical}, other quality issues are analyzed in \citet{ettinger2020bert, jones2022capturing}.}. Nevertheless, many models who almost certainly have such defects are being deployed as live products, and this is likely to overall bias the internet commons towards such events.
        \item Many people are trying to use text models as knowledge bases, querying them for answers much like Google or Wikipedia, which makes the above more problematic. The input data is not sanitized for truthfulness (that is near impossible at the scale at which data is collected), therefore there is in fact no guarantee that these models will say the correct answers. However, the attempts at using them in these applications will lead to mistakes of fact, many of which will not be caught. Meta’s Galactica language model was trained to generate scientific wiki essays and has notoriously emitted many subtle but confident-seeming scientific falsehoods. This might lead to a proliferation of misinformation, either accidentally or purposefully.
    \end{enumerate}
    \item \textit{Accessibility and generalizability.} GFMs are hosted publicly on websites such as HuggingFace and Github, and require comparatively little technical knowledge to use, making them very accessible. They are also more accessible and generally applicable than previous methods of autogenerating content, such as articles that are optimized for ranking well on Google using automated techniques like Markov chains, or scraping RSS feeds, or automated synonymizing \citep{google_spam_agc, auto_generated}. Compared to other algorithms applied for content generation, GFMs require less specialized development for particular use-cases, and thus auto-generation can proliferate in far more domains than previous techniques.
\end{enumerate}

Such issues may greatly degrade the quality of the information commons and require some level of restructuring of the internet e.g. intensifying and requiring new solutions to the problem of how we detect, filter and rank machine- vs human-generated content.

\textbf{It is difficult to determine the criteria for what generations are desirable, and to detect and counter undesirable generations.} There are wide disagreements over the extent and criteria of what generations should be allowed or not \citep{mcgee2023chat}. Furthermore, even if there was agreement, it is difficult to discover and mitigate all harms. Many generations are wrong or otherwise undesirable in subtle ways, as stated above, or ways that vary depending on context, making the creation of classifiers and content filters a hard task. Furthermore, the technical problems of permanently tagging/stamping data as machine-generated vs. human-generated is difficult, and there may be race dynamics between those developing methods to detect AI-generated content and those trying to outwit detectors \citep{yu2022responsible}. This contributes to the difficulty of ensuring safety.

\textbf{These generated outputs start to become part of the information commons.} Conversational response and code auto-completion are common uses of GFMs among other products \cite{zhang-etal-2020-dialogpt}, students are already using GPT-3 tools to write convincing school essays \citep{ai_school}, and people can automate the creation of mis/dis-information e.g. via fake news generation, fake product reviews, and spamming/phishing \citep{buchanan2021truth, zellers2020defending, adelani2019generating}. Given the many undesirable properties of generated outputs, this might “pollute” Internet-based datasets, including training for future generative models.

The outputs of state-of-the-art GFMs tend to be much more indistinguishable to that of humans than previous algorithms \citep{weiss2019deepfake}. It will become increasingly difficult to tell what is AI-generated vs. human-generated content, and to filter one from the other. Countermeasures for distinguishing GFM outputs could be developed, as could new attacks that overcome those countermeasures, but it is likely that much GFM-created content will join the information commons as plausibly human-created content.

Issues of training data quality are already cropping up with low-resource languages on Wikipedia, where content in Scots turns out to be patently wrong \citep{scots}, or most Cebuano and Swedish articles turn out to be bot-generated \citep{wiki_languages}. Researchers generally take Wikipedia to be a good source of high-quality human-written content, and hence train AI on this low-quality, bot-generated content, resulting in models with lower quality output \citep{Kreutzer_2022}.

If GFM outputs join the commons and become part of future datasets for newer GFMs, this may bias newer models towards older, established patterns and make future quality improvements more difficult, especially if input datasets are not carefully filtered or utilized \footnote{An executive at code completion company Tabnine mentions this concern in an interview \citep{tabnine_interview}.}.

This may be like the effect of nuclear weapons and climate change on radiocarbon dating \citep{radiocarbon, jones2022carbon}, where samples after a certain time cannot be used, with standard techniques, to accurately date the material. After a certain time, the information commons may not be high enough quality for standard ML training techniques to work.

This may make it even harder for researchers, such as computational social scientists, to work reliably on Internet-based datasets. Available datasets already greatly limit such research \citep{gaffney2018caveat}, and the large-scale creation of machine-generated content will add new challenges.

GFMs may erode self-determination and democracy. In many ways, GFMs could be used against self-determination, democracy, and even by state or semi-state actors in warfare operations e.g. by being used for personalized persuasion or disinformation. Genuine popular movements could also be discredited by being accused of being composed of GFM-generated material. Additionally, the content filters and other design choices are generally determined by a small team with little participation or diverse input which arguably creates an insufficiently democratic situation, especially as GFMs pertain to issues that impact political discussion (e.g. Meta’s Galactica AI filtered out legitimate scientific research to do with race or AIDs \citep{heaven2022meta}). Some of these problems may become less salient given the trend of open-sourcing models.

GFMs may disincentivize contributions to the digital commons. People may stop writing and drawing in favor of using generative models, decreasing the production rate of new human-written material. (On the other hand, more people might engage with creating material with the aid of GFMs, with still significant human input, or with greater overall creative output.) Additionally, people may not want to release their non-GFM-enabled creations into the commons as much. Incentives to, for example, open-source one’s code or publish high-definition versions of one’s artwork may decline, because of fears of labor replacement or dislike of adding to training data e.g. not getting attribution or publicity, feelings of misalignment with the purpose of GFMs.

It may also decrease incentives for artists to make their work legible to algorithms (e.g. artist Greg Rutkowski’s work has been popular for use in GFMs, because he adds alt-text to his images \citep{greg}), which is bad for data science and ML, accessibility, and for having a free and open art commons. More may be hosted on private, non-scrapable services and less completely publicly, which gives more power to companies that wall off information (e.g. Facebook content cannot be scraped and used in training sets; people may publish there instead of on more open platforms like Artstation). This closes the digital commons off further. Overall, the digital commons may receive a large amount of lower quality AI-generated content, whereas the release of human-generated content is disincentivized.

\section{Existing proposals and their pros and cons}

Although the space is nascent, several proposals have been put forward to mitigate potential pitfalls of GFMs on societal outcomes. They each have pros and cons regarding their ability to address systemic risks. These include:

\renewcommand{\arraystretch}{1.5}
\begin{tabularx}{\textwidth} { 
  | >{\raggedright\arraybackslash}X 
  | >{\centering\arraybackslash}X
  | >{\centering\arraybackslash}X 
  | >{\centering\arraybackslash}X 
  | >{\raggedleft\arraybackslash}X | }
 \hline
   & \textbf{Data dividends/taxes} & \textbf{Retraining for automated sectors} & \textbf{Stricter copyright and licensing} & \textbf{Individual data rights and ownership} \\
 \hline
 \textbf{Risk(s) addressed}  & Economic concentration, labor automation (depending on use of funds)  & Economic concentration, Labor automation & Poisoning information sphere (depending on success of copyright), misaligning contribution incentives & Economic concentration, misaligning contribution incentives \\
 
 \textbf{Actor}  & Government (potentially in multiple jurisdictions), GFM-building or utilizing companies & GFM-building or utilizing companies (if voluntary contribution); Government (if regulated) & Copyright / licensing bodies (ex. Creative Commons), Courts (if lawsuits succeed), Public sector & Government (through regulation), standards groups, content platforms \\
 \textbf{Beneficiary}  & Citizens, through the pipeline of tax → public sector entity → social safety net / public goods & Displaced workers and future workers & Holders of copyrighted or copyright-able data, potentially large rights-holding organizations over small & Individual data creators \\
\hline
\end{tabularx}

Most of these have implications for the sustainability of the digital commons, but do not take into account many of the original risks listed. Pros and cons for each are listed below to explicate the reasoning behind each idea and to show where they have overlaps with the more commons-focused proposals in the next section.

\subsection{Data dividend-style taxation approaches:}

A data dividend has been proposed for other issues in the past \citep{feygin2021data}; it is aimed at large private entities hoarding user data. In this case it may not be as relevant.

Pros
\begin{enumerate}
    \item It is easier to apply laws on the company compliance level rather than e.g. requesting individual users to opt-in to training sets.
    \item This is an avenue for redirecting money back to the commons, by funding public goods, digital or otherwise. (The original data dividend proposal, created for the state of California, suggested funding e.g. baby bonds, or education.)
\end{enumerate}

Cons
\begin{enumerate}
    \item There are downsides to people thinking of their data primarily in terms of monetary value rather than as a right \citep{tsukayama2020getting}.
    It may be difficult to determine how and what to tax; should taxes be incurred by the entity that trains the GFM, or entities that deploy products that use GFMs (even if they did not train the GFM)? Should taxes be levied only on particularly high-revenue companies? Should taxes be levied on the basis of the number of unique data-points used in the model, and how is that counted? There are various potential distortions in incentives that need to be considered.
    Taxes must be levied by particular jurisdictions, but the issue may be faced in other jurisdictions, or properly solved only at an inter-jurisdictional level. There may be legal issues with enacting taxes at the state / local level, as recently seen with the Maryland Digital Advertising Tax \footnote{The law faced two issues: 1) state/local interference in interstate commerce (constitutionally within Congress's jurisdiction), and 2) IFTFA preemption of state/local law. \citep{internet_tax_freedom_act}}.
\end{enumerate}

\subsection{Directly supporting automated sectors with training or pay:}
Pros
\begin{enumerate}
    \item If primarily concerned with labor automation, may be an easy intervention to immediately address short-term symptoms. 
\end{enumerate}

Cons
\begin{enumerate}
    \item It’s difficult to provide a future-proof service that will train the right people in the right ways.
    \item Broader economic reforms and welfare initiatives may be more effective.
\end{enumerate}

\subsection{Copyright and licensing approach:}

Calls for a stricter application of fair use doctrine or otherwise relying on copyright law have been made, which may require companies to license images, text or code snippets from individuals or companies e.g. Shutterstock. However, it is unlikely that copyright is a sufficient data regulation regime to address concerns. A new regulatory framework is likely to be needed.

Pros
\begin{enumerate}
    \item Leverages an existing framework and tools. 
\end{enumerate}

Cons
\begin{enumerate}
    \item It is likely that most AI training is fair use, given that it is usually sufficiently transformative to be different from source material. Some, notably the head of Creative Commons, have said that even facial recognition model training is fair use \citep{merkley2019use}. But expression is not the same as recognition, and especially for artists whose expressions are being copied (e.g. prompts to image models to generate images in the style of some artist) the fair use case may not hold up \citep{fair_learning}).
    \item A heavy-handed copyright approach may overly narrow the fair use doctrine and prevent many good applications being created.
    \item This may boost the advantages of those with proprietary data over e.g. smaller companies training on large, Internet-collected datasets with heterogeneous copyright licenses, and/or lead to more data-hoarding by private platforms.
    \item May prompt large rightsholder organizations to change boilerplate terms to acquire training rights to everything they contract for.
\end{enumerate}

\subsection{Advancing an individual data rights and ownership approach: }
Similar to the copyright and licensing approach, but this is a more drastic approach that has been advocated where companies pay individuals for the rights to their data.

Pros
\begin{enumerate}
    \item Clear tie-in between value captured from individuals and value returned to users. 
\end{enumerate}

Cons
\begin{enumerate}
    \item Will be difficult to implement in practice for public web data. Much of this data is difficult to trace to particular individuals or groups that can be said to own it.
    \item It would greatly change AI as an industry if done for public web data. It may transform the kind of data used and thus the AI capabilities developed; web-crawled data is convenient for developers right now given lack of regulatory overhead, but will become less popular if overhead increases.
    \item It would disproportionately benefit those who provide more data to algorithms, which may lead to undesirable incentives for individuals to e.g. be “more online”.
    \item Does not approach the problem on a collective or commons-oriented level, which is where it originates, and thus does not account for effects on general quality of information commons.
    \item May prompt large rightsholder organizations to change boilerplate terms to acquire training rights to everything they contract for, much like the above copyright and licensing approach, thus becoming largely ineffective.
\end{enumerate}

\section{Proposals to evaluate: consortia for monitoring and standards, data contributions to the commons, input-data-based governance}

The below proposals (each column) have been developed with the risks in Section II in mind.

\begin{tabularx}{\textwidth} { 
  | >{\raggedright\arraybackslash}X 
  | >{\centering\arraybackslash}X 
  | >{\centering\arraybackslash}X 
  | >{\raggedleft\arraybackslash}X | }
 \hline
   & \textbf{Consortia for monitoring, auditing and standards-setting} & \textbf{Norms or rules for GFM companies to contribute high quality data to the commons} & \textbf{Governance structures based on input data to model training} \\
 \hline
 \textbf{Risk(s) addressed}  & Poisoning the information sphere, eroding democracy.
Can provide greater insight into the reality of other risks, incl. homogenizing content, labor impacts, and economic concentration
  & Poisoning the information sphere
Can provide greater insight into homogenizing content, labor impacts
 & Eroding democracy, driving economic concentration, labor automation, misaligned contribution incentives \\
 
 \textbf{Actor}  & GFM-building and utilizing companies (if voluntary; most likely); Government (if regulation; unlikely) & GFM-building and utilizing companies; Researchers & GFM-building and utilizing companies; Individuals, groups, and organizations with relevant data and expertise; eventually perhaps government (for necessary regulation around data transfer and protection + fiduciary responsibilities \\
 \textbf{Beneficiary}  & GFM companies (standards-setting is likely to be positive-sum); Researchers (better access to data); users of GFMs/Internet & Researchers (better access to data); users of GFMs/Internet & Individuals, groups, and organizations with relevant data and expertise; GFM companies (access to higher-quality data from valuable domains, hard to scrape) \\
\hline
\end{tabularx}

\subsection{Proposal I: Consortia for monitoring, auditing, and standards-setting.}

Solutions to commons-based problems that rely on the action of individual stakeholders are historically unlikely to work. However, consortia-based solutions, in which stakeholders come together and form collective institutions to provide or fund shared goods necessary to maintain the commons, have been robustly helpful in many domains.

Examples of this include standards-setting bodies, peer review protocols and trade associations. Furthermore, given that GFMs are an early technology, establishing a GFM-governing consortium that develops strategies and agrees on new policies, will help establish best practices for responding to emerging risks or opportunities of future ML development (much like the W3C, which was established early in the Internet’s lifetime). This may look like the Coalition for Content Provenance and Authenticity (\href{https://c2pa.org/}{C2PA}), which focuses on content provenance and authenticity certification standards for media.

The consortium could connect the ecosystem in productive ways, e.g. connecting companies that train and release open-source GFMs with companies that deploy application-specific uses of GFMs. They could also be specific to different parts of the GFM ‘pipeline’ e.g. a consortium that focuses on issues specific to GFM application-deploying companies, rather than issues pertaining to  companies that train them.

A working solution, however, must also avoid standards-capture by incumbents. Among other requirements, this means that those in charge of setting the standards or policies should not all be from GFM companies, and should be broadly representative of different interests.

There are two sides to this proposal:

\subsubsection{Resourcing}

\textbf{Voluntary memberships and subscriptions.} Much like private organizations contribute to standards-setting for the internet, GFM companies may contribute time and money to collective governance structures due to useful outcomes and good public perception. This is the model pursued by organizations such as the Linguistic Data Consortium and the Partnership on AI.

\textbf{Regulation and taxation.} If the above does not hold, we might want to explore public money, grants or even data-dividend based models for funding such a body. A public funding model is pursued by organizations such as the National Consortium of Intelligent Medical Imaging in the UK. 

\subsubsection{Organizational duties:}

\textbf{Monitoring.} We cannot know what we don’t track, and understanding the risks and opportunities of GFMs will require monitoring and analysis, e.g. setting up data science pipelines to monitor whether they are causing the spreading of misinformation. Researchers have pointed out that we cannot understand systematic harms related to how social media affects society without better monitoring structures in place to collect high-quality data about its effects — the situation is arguably worse in generative AI, which is very new, and where research on information ecosystem impacts is nonexistent \citep{lubin2022socialmediapolluting}. We suggest efforts at identifying and potentially sponsoring such research. 

\textbf{Auditing.} Consortium-based organizations are a natural choice for third-party AI audits, which have been called for by civil society groups, researchers, and even private entities. They are also a possible avenue for setting norms and standards around internal or voluntary audits.

\begin{itemize}
    \item These audits can implement and modify existing proposed frameworks for audits such as SMACTR for internal auditing \citep{raji2020closing}, creating audit trails (including documenting specific design and data choices, as well as tests), and documenting organizational processes and accountability structures.
    \item Audits can look for problematic behaviours of discrimination, distortion, exploitation and (likely less applicable to generative rather than classification algorithms) misjudgement, as well as create robust and transparent suites of metrics and establish compelling baselines for behaviour (as noted by \citet{bandy2021problematic}).
    \item Given that a model can be trained/fine-tuned and deployed by different entities, audits can happen at the level of the model or the application (which may involve different organizations).
\end{itemize}
Note that auditing could be a separate proposal on its own outside of the consortia context, as it is possible for governments, for-profit and non-profit auditors, or some international organisation to audit. However, a key limitation of AI auditing in general is the lack of an industry ecosystem for doing so, so bundling this under the responsibilities of a more general consortium in order to create a needed home and mandate for auditing, may make sense.

\textbf{Standards for auditing and transparency.} Proposals around algorithmic audits (both processes and metrics), data cards, and model cards for AI in general are gaining traction. However, release and use are not standardized, and datasets on which widely used models are trained are still poorly understood. Without this understanding, both impacts and contributions to the commons cannot be appropriately tabulated. Standards-setting bodies, either for AI at large or GFMs in particular,  can determine appropriate venues and forms of information release and enable industry-wide transparency and accountability.

\textbf{Developing shared tools.} Tools to filter for high-quality data and/or to detect AI-created data will be needed to combat the problems of digital commons pollution. It is not clear if access to such tools will be equitable as they are likely to be developed by the best-resourced companies. For more equitable and widespread development of and access to these tools, consortia-based development of them may be helpful, similar to the approach of the C2PA.

\subsection{Proposal II: Norms or rules for GFM companies to contribute high quality data to the commons.}

GFM companies could create, as a norm or a rule, gold standard datasets usable by other entities. These datasets could be the datasets they train models on, or datasets related to how models are deployed or used by other parties. Numerous companies creating GFMs do not disclose their datasets or their dataset-cleaning methods, making it difficult to understand the properties of the resulting models or for others to replicate work. In general, information release related to GFMs and how they are deployed is greatly lacking.

Companies could be encouraged to admit data on deployment to a privacy-preserving but accessible repository through a body like NIST, whenever they release a model for use. This would let researchers look at data for different deployments over time, and determine e.g. labour impacts of GFM deployment. They could also be encouraged to admit high quality datasets they use to improve the model, e.g. one with high quality labels of harmful GFM use cases, which will enable other companies to more easily adhere to standards of safe model deployment. Companies would need to follow specified scientific data collection procedures and be transparent about how they sample and process data. There would need to be external scrutiny to ensure the validity of collected data.

Overall, this would decrease “moats” and increase competition, increase transparency and researcher access to understanding what data deployed models are trained on, ensure safer deployment, and also generate useful metadata for tracking labor impacts and economic effects of models.

\subsection{Proposal III: Governance structures based on input data to model training.}

High quality data is a key bottleneck for training AI, now and in the future, so data governance is a key lever for governing generative models. Data cleaning and processing is critical to training a performant GFM. Another method to ensure high quality data is to focus on actively generating human feedback data to improve models, such as reinforcement learning with human feedback \citep{stiennon2022learning, christiano2023deep}, We propose that this could be done in return for ownership stake or governance rights. 

Multiple models are possible:
\begin{enumerate}
    \item Human feedback is provided by users to improve the model, in return for stake. (Sam Altman has expressed interest in giving “credits” to those who are giving feedback on utterances and thereby improving ChatGPT; other forms of feedback or stake could be explored here and in other contexts, e.g. direct feedback on content filters beyond feedback on specific utterances, some form of “shares” in the model beyond credits for using the model.)
    \item Individuals or groups with specific expertise could contribute fine-tuning data. Better data provenance, in this way, could enable easier tracking of truth and citations, and thereby combat model hallucinations. 
    \item Organizations with highly private data that is necessary for valuable applications (ex. financial fraud detection and medical diagnosis) could act as fiduciary data intermediaries for inputting that data into models via privacy-aware training approaches. Benefits could be distributed to organizations (to support fiduciary capacity) or back to communities. 
\end{enumerate}
The details of this require further research and comments are welcome.

For GFM applications that require specific data from a clearly-defined constituency (e.g. training or fine-tuning an image generation model on specific artists’ work), non-profit organizations that act as data trusts can be set up to help data-owners interface with companies that train models.

Groups that have specific use cases for GFMs may also come together to collectively create and govern GFMs together themselves, thus governing both data and model. For example, a specific group of artists may come together to jointly train and decide governance of an image model that they are able to monetize themselves.

Early individual case studies of the above structures may provide excellent input to policymaking around feasible data and model governance structures for the broader public.

\subsubsection*{Acknowledgments}
An enormous thank you to (in alphabetical order) Miles Brundage, Jack Clark, Cory Doctorow, Yakov Feygin, Ben Garfinkel, Yacine Jernite, Katya Klinova, Pamela Mishkin, Aviv Ovadya, Sandy Pentland, Irene Solaiman, Nick Vincent, and Glen Weyl for their immensely helpful comments and critiques on this paper. In addition, we'd like to thank many members of the Center for the Governance of AI who gave us feedback on this during a Work-in-Progress session.

\bibliography{iclr2021_conference}
\bibliographystyle{iclr2021_conference}

\end{document}